\documentclass[letterpaper,doc,notimes]{apa6}
\usepackage[normalem]{ulem}
\usepackage[english]{babel}
\usepackage[utf8x]{inputenc}
\usepackage[numbers]{natbib}
\usepackage{graphicx}
\usepackage{amssymb}

\title{IRR: Review Paper}
\shorttitle{Detecting Code Clones: A review}

\begin{document}

\begin{center}
    \Large
    \textbf{Detecting Code Clones: A review}
    
    \vspace{0.2cm}
    \large
    Informatics Research Review (IRR)\\
    University Of Edinburgh
    
    \vspace{0.2cm}
    \textbf{Ogechi Onuoha (s1459203)}
    
    \vspace{0.9cm}
    \textbf{Abstract\\}
\end{center}
Code clone detection is involved with detecting duplicated fragments of code within a code base. Detecting these clones is useful for maintenance operations which require editing the clones. The tools developed are expected to be robust enough to identify clones even when they have been modified, whilst preserving reasonable recall and precision rates.  It is also expected that these tools be easily adaptable to different programming languages. The major approaches to this problem has involve the use of direct string matching, token comparison or comparison using abstract syntax trees. It is difficult to compare detection tools due to the absence of a standardised framework for measurement. More work should be done to make the existing tools useful for other practical/industrial purposes.

\vspace{1.2cm}

\section{Introduction}
Duplication of code occurs when a software programmer copies a portion of working code to another location either within the same code base or to another code base. Usually these code fragments are edited slightly to fit the use case in the new location. As duplication is not included in the documentation of the application, this can lead to complications when managing the software; especially when the copied code contains bugs within itself. Hence, when such a code fragment needs to be edited, usually for the sake of bug fixing or to implement an update, all duplicates of that code will require a corresponding adjustment to prevent bugs and inconsistencies in the final application. For such a maintenance operation to be complete, all the copies of the code fragment need to be identified prior to modification. In a large code base, this becomes nearly impossible to achieve manually. Code clone detection algorithms and tools have been developed to tackle this challenge. 
Recent software development platforms have a refactorisation feature which does a top level editing. However, this is not sufficient for detecting or editing code clones especially when portions of the clones have been modified.\\

Defining what constitutes a code clone may be difficult, but generally speaking, code clones are those duplicates in a code base that pose copy-paste problems as defined above. Hence, programming language constructs, idioms and other such recurring statements should not normally be considered as duplicated code, since they do not pose such copy-and-paste
problems\cite{ducasse2006effectiveness}. Many code detection approaches require parsing of the source code and using a pattern recognition or data mining technique to fetch the clones.\\

There are two ways to declare two fragments of code as duplicates. One way is to detect similarity in the text that constitutes the fragments, another method is to investigate similarity in function i.e. investigating if the input and output of the code fragments are similar.
Intuitively, high similarity in the text that constitutes the  source code is an indication of similarity in semantics or resulting functionality, regardless of if variable names are changed.
The validity of the notion that code clone duplication negatively affects the quality of code was investigated by \citeauthor{hordijk2009review} \cite{hordijk2009review}. Since making a copy of a tested (legacy) code fragment is less intrusive and faster than rewriting the code from scratch. This is even more so if the code base is not modular. 
A lot of effort has gone into the investigation of code clones. Similar research efforts have gone into related fields such as code clone evolution, reasons why code clones occur, or the impact of clones on software quality. Code clone evolution is the study of how code clones evolve from one version of source code to the next \cite{pate2013clone}. Such a study requires the prior detection of the clones as well as version tracking.

\section{Approaches to code clone detection}
The approaches to code clone detection, as observed in literature, can be categorised into four main headings.
\begin{itemize}
\item Text or string based approach. This method is the easiest to implement as it requires comparison of sequences of strings for similarity\cite{baker1993program}.
This method is not robust to spaces and minor modifications except advanced pre-processing operations are applied.  \citeauthor{ducasse2006effectiveness}\cite{ducasse2006effectiveness} found that string based approaches have high recall, which is the fraction of detected clones that are actual clones, and precision which is the fraction of detected clones to all proposed candidate clones, (See figure 1). String based approaches can be easily adapted to new languages. An advantage of using the string based approach is that the researcher can leverage the optimised algorithms that exist for string comparison e.g. the suffix tree. It is also cheap as it excludes a need to acquire or develop a specific language parser which is required for other approaches. The characteristic work using this method is Dup\cite{baker1995finding} by \citeauthor{baker1995finding}.

\item Lexical or Token based approach. This approach is similar to the string based approach since it requires a line based comparison. The significant difference being it lexes the strings using a language dependent parser and uses the generated token sequences for comparison.
 This method is considered heavyweight \cite{ducasse2006effectiveness} since it requires the acquisition of a language parser which is specific to the language and dialect used in development. CCFinder\cite{kamiya2002ccfinder} and CPMiner\cite{li2004cp} employ this approach; although CCFinder only detects contiguous token strings while CPMiner is able to detect non contiguous token sequences\cite{kim2005empirical}. 
\item Tree based approach. This method uses abstract syntax trees (AST) which are built using tokens generated after line based parsing. The subtrees of the resulting AST may be converted to a hash using a hash function that ensures that similar trees are assigned the same hash value. Alternatively, a fingerprinting function may be applied to the AST before hashing. The resulting hashes are then clustered using a clustering algorithm. Local Sensitivity hashing (LSH) was employed by \citeauthor{jiang2007deckard}\cite{jiang2007deckard} to perform clustering after fingerprinting to ensure efficiency. Subtrees in the same cluster are clone candidates. 

\item Semantic approach. Semantic clones are clones detected based on similarity in their observable behaviour or functions \cite{koschke2007survey}.These types of clones are the most difficult to detect. Many times this approach does not scale to large code bases.  
\end{itemize}

\begin{center}
\includegraphics[width=0.75\columnwidth]{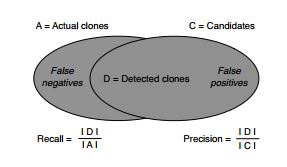}\\
\end{center}

\begin{center}Figure 1: Recall and precision\cite{ducasse2006effectiveness}\end{center}

The above methods are not robust to advanced plagiarism, if the programmer is  experienced, he/she may be able to edit a program enough to avoid detection by the above methods \cite{koschke2007survey}. However, the order of a cloned code usually remains the same, as there is a limit to the reordering of a code that preserves its semantic meaning.

It is worth mentioning that some research has been done to attempt combining text based approaches with semantic comparisons as seen in the work of \citeauthor{leitao2004detection} \cite{leitao2004detection}.

\section{Types of code clones}
The types of clones can be grouped into 3\cite{basit2005detecting}\cite{wahler2004clone}:\\
Type 1 clones (Exact clones) - where a clone is a clear copy of a code fragment with very minor editions such as deletion of white space or addition of comments.\\
Type 2 clones (Parameterised clones) - where the cloned code is modified by changing variable and method names, usually in an attempt to mask it from detection or to make it better fitted for its destination. Code fragments such as $"sum = sum + 5"$ and $"total = total + 5"$ are considered as Type 2 clones.\\
Type 3 clones(Gapped clones) - where whole statements are deleted or relocated within the clone. These are the hardest to detect.

Most of the tools available are able to handle clones of Type 1 with a few doing well with clones of Type 2. Detecting Type 3 clones was relatively successful in the work of \citeauthor{krinke2001identifying} \cite{krinke2001identifying} although it compromised on its ability to detect Types 1 and 2 clones \cite{wahler2004clone}.

A distinction has also been made between simple clones- clones which use copied portions of code and structural clones which are based on "mental templates" which programmers use repeatedly to solve the same problem \cite{basit2005detecting}. An example is the method of implementing a swap of two variables which remains unchanged in many applications.

\section{Challenges facing code clone detection}
Code clone detection is akin to searching for people with the same features in a population. The search space is in an order of $O(n^2)$\cite{ducasse1999language} where n is the number of lines in the source code after pre-processing.
It is obvious that manual method of identifying clones in a code base becomes inefficient as the code base grows. A good clone detection tool must be able to cope with large code bases. This is called scalability.\\ 
While automating the process of detection, Questions like "How similar should two code fragments be to be declared clones?" are not uncommon. A useful detection tool must have a clear definition of its basis for declaring clone candidates. This definition is not fixed and it varies from application to application depending on the judgement of the researcher. Unfortunately, this  affects the quality of the clone candidates.

Also, the compromise required between recall and precision poses a challenge. As explained by \citeauthor{ducasse2006effectiveness},using the notation in Figure 1, if a detection tool presents C code fragments as candidate clones, we aim to bring C as close as possible to the actual clones A, contained in the code base. If D is the set of code fragments in C that are accepted as actual clones, then Figure 1 defines Recall and Precision. The problem of recall and precision is a problem of false positives and false negatives.
False positives occur when the detection tool wrongly labels code fragments as clones. This negatively impacts precision. Similarly, false negatives occur when the detection tool fails to identify actual clones as clones. This will occur if the programmer is sufficiently experienced and is able to modify the code sufficiently to mask the clones. This affects recall. Over these parameters, we seek a system that shows a good level of recall as well as precision. Unfortunately, techniques
that have high recall (detect many clones) return many false positives (lower precision). The converse is also true. However, many times the compromise between precision and recall is biased towards recall.

Finally, code clone detection is heavily dependent on the specific language and dialect. Thus, such a tool created for one language may require an expert technologist to adapt for detection in source code written in another language. This is especially true when a more advanced approach is used such as the tree based or semantics based approaches.

\section{General steps used in code clone detection}
Clone detection usually involves three major steps. A pre-processing stage which takes away uninteresting content like comments and white space. Afterwards, the source code is transformed into a compatible
format e.g strings, tokens, trees or even vectors. Then, a comparison
algorithm is then applied to the resulting data\cite{ducasse1999language}.

To detect similar code fragments or clones, a researcher must define 'similar' through a similarity function. This is a function or algorithm that is used to compute the degree of nearness the clones have to each other. Also, a threshold (T) must be defined which gives a minimum value, usually of a distance measure e.g euclidean distance, below which code fragments can be declared as clones\cite{baker1993program}. This is particularly useful if comparison is done using clustering.
Hence, code fragments $C = {c_1, c_2,...,c_n} $ are said to be clones if $s(c_1, c_2,...,c_n) \geq T$ \cite{pate2013clone}. Two code fragments which are clones of each other are said to be clone pairs and three or more clones form a  clone group or clone set or clone class \cite{pate2013clone}\cite{hordijk2009review}. Intuitively each clone in a clone set should form a clone pair with every other clone in a clone set.

The following methods were found as underlying steps followed in most code clone detection endeavours.

1. Pre-processing or de-noising: This step prepares the code base for analysis. It removes the elements that are unnecessary or unwanted in the detection process. These usually include, but are not limited to, white spaces and comments, which do not contribute to the definition of a clone. Block delimiters, and pre-processor directives are also usually removed. In languages that are not case sensitive, the resulting code base is converted to lowercase\cite{ducasse2006effectiveness}. Depending on the method to be used for comparison, extended processing may be necessary. E.g. for token based methods, the code units are tokenised using the specific language parser for the code base.

2. Generate fragments for testing: This is a sequence of code units generated in the previous steps. Usually the fragments are of a length defined by the researcher and implemented in the tool as a 'window'. This window defines the length of the shortest code fragment into which the code base will be broken. This window must be small enough to capture relevant units but not too small to prevent identifying irrelevant, uninteresting pairs as clones. Some detection tools allow for overlap of code fragments.

\begin{center}
\includegraphics[width=0.75\columnwidth]{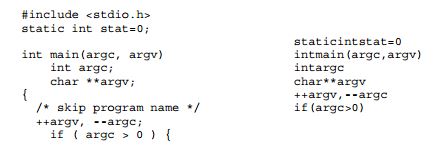}\\
\end{center}

\begin{center}Figure 2: Results after pre-processing \cite{ducasse2006effectiveness}\end{center}

3. Comparison: Comparison of the units using a similarity rule defined by the researcher and designed with a specific objective i.e syntax comparison or semantics comparison. Comparison may be done by direct string matching or clustering. 

4. Analysis of results. This usually  involves measuring recall and precision of the tool. It would involve summary statistics of the clones found as well as presenting and comparing the results  to the results obtained by other researchers.

\section{Techniques used for pattern recognition in clone detection tools}
Various approaches have been used to achieve Pattern recognition in code clone detection. 

Pure string matching algorithms have the advantage of being language independent and easy to configure or modify.
Some tools that use the string based approach, are based on Suffix trees which are compressed trie data structures that allow for efficient searching of strings. A suffix tree can be built in time O(N) using O(P) \cite{manber1993suffix} time for a query. Here, N is the length of the string to be used for tree construction and P is the length  of the substring to be searched. In the case of Dup, a new data structure called parameterised suffix tree or p-suffix tree was developed. Such a tree is constructed using "parameterised strings"\cite{baker1995finding} which keep track the position of the substrings as well as the substrings themselves. For every match, Dup outputs the number of matching  lines, the 
matching intervals, and a list of the nonidentical parameters for each p-match\cite{baker1995finding}. Token based tools, such as CCFinder and CPMiner, use tokens after language specific parsing. They are known to be more sensitive to modifications but have higher recall. Tree based approaches make use of tokens and are capable of detecting modified clones. On top of the AST, \citeauthor{wahler2004clone} created an internal representation using XML and applied the frequent itemset data mining technique which attempts to find items which are subsets of a larger collection\cite{wahler2004clone}.\\
Metric based pattern recognition makes use of fingerprint functions instead of directly comparing code fragments. It is based on the notion that source code can be converted to numbers or numeric values and in some tools, such as Deckerd, these numbers may be built into vectors. This method usually builds on the AST generated from the source, subsequently applying the fingerprinting function to convert the sub trees to their numeric or vector forms. Vectors are then clustered and subtrees in the same cluster are declared as similar. 
Program dependency graphs (PDG) have been suggested to detect semantic based clones. PDGs hold information about the semantics of the source code\cite{bruntink2005use}. The approach takes the semantics of the source code into consideration while clustering similar codes together \cite{komondoor2001using}. This approach is able to identify "non-contiguous, reordered, and intertwined clones, and the likelihood that the clones that are found are good candidates"\cite{komondoor2001using}.
Efforts have also been made to improve clone quality by combining two methods or tools as seen in the work of \citeauthor{basit2005detecting} \cite{basit2005detecting}.

\section{Comparison of detectors and clones}
Comparing detectors is done based on the concepts of recall and precision.   
Code clone detectors are hard to compare especially because the motivation behind each detector is different. Such variations include \cite{hordijk2009review}:
\begin{itemize}
\item Different approaches employed in the abstraction of the source code
\item Different target programming languages
\item Different definitions and parameters 
\item Different goals with which tools were designed or research was performed.
\end{itemize}

This also directly affects the comparison of the clones detected using these methods because
\begin{itemize}
\item The tools use different methods for generating the comparable units and these units are not easily convertible to each other.
\item Different tools use different thresholds. A single clone detected by one tool may be detected as two separate clones in another\cite{hordijk2009review}. This makes one unable to objectively compare the number of clones detected by one tool to those of another tool even when the code base is the same.
\item The clustering methods or similarity functions are different \cite{hordijk2009review}
\end{itemize}

\section{Further work}
Code clone detection tools are yet to be used fully for non research purposes \cite{hordijk2009review}
Practically, Code clone detection evaluation may be investigated for various reasons. One may want a measure of the maintainability of a code base, or to evaluate the evolution of the code base, for plagiarism detection or even to evaluate the ease of code re-factoring \cite{hordijk2009review}. Many of the tools developed are not developed to optimise one or another of these applications, nor are they designed to evaluate the properties of the clones detected. This makes it difficult for practical/objective use of the available tools to these ends. More industry applicable features are needed in the tools such as is found in Dup which tries to give an estimate of how a code base could have been improved if clones were eliminated during the development and replaced by such things as macros or procedures. Here, the assumption is if n clones are found over a particular code fragment then n-1 clones could have been avoided. \cite{hordijk2009review}
Also, having highlighted the difficulty in comparing code clones, \citeauthor{hordijk2009review} identified the need for a unifying framework over which the code clone detection parameters can be defined. 

\section{Conclusion}
Although the real effect of code clones have been argued, there is no doubt that bug ridden code clones increases the time and cost implications of code maintenance operations. Comparisons of the tools developed so far, however difficult, shows that there is no tool or method with returns better results in all situations, although the researchers often claim so \cite{rysselberghe2004evaluating}. Although a fair compromise may be made, recall and precision are the key aims of every code clone detection tool. Type 1 and Type 2 clones have been detected with relative ease. However, a lot of work still needs to be done in detecting Type 3 clones.

\newpage
\bibliographystyle{plainnat}
\nocite{*}
\bibliography{ref}{}

\end{document}